\documentclass{iau}
\usepackage{graphicx}

\title[JD 11.~~Caustics and Large-Scale Structure] 
{Statistics of Caustics\\ in Large-Scale Structure Formation}

\author[Feldbrugge et al.]   
{Job L. Feldbrugge$^1$, Johan Hidding$^1$, Rien van de Weygaert$^1$}

\affiliation{$^1$Kapteyn Instituut, University of Groningen, \\ Postbus 800, NL-9700AD, Groningen, the Netherlands \\ email: feldbrug@astro.rug.nl }

\pubyear{2014}
\volume{308}  
\pagerange{119--126}
\jname{The Zeldovich Universe: Genesis and Growth of the Cosmic Web}
\editors{R. van de Weygaert, S.F. Shandarin, E. Saar \& J. Einasto, eds.}

\begin{document}

\maketitle

\begin{abstract}
The cosmic web is a complex spatial pattern of walls, filaments, cluster nodes and underdense void regions. It emerged through gravitational amplification from the Gaussian primordial density field. Here we infer analytical expressions for the spatial statistics of caustics in the evolving large-scale mass distribution. In our analysis, following the quasi-linear Zel'dovich formalism and confined to the 1D and 2D situation, we compute number density and correlation properties of caustics in cosmic density fields that evolve from Gaussian primordial conditions. The analysis can be straightforwardly extended to the 3D situation. We moreover, are currently extending the approach to the non-linear regime of structure formation by including higher order Lagrangian approximations and Lagrangian effective field theory.

\end{abstract}

\firstsection 
\section{Introduction}
\noindent The large-scale structure of the universe contains clusters, filaments and walls. This cosmic web owes its intricate structure to the density fluctuations of the very early universe, as observed in the cosmic microwave background radiation field. For discussions on the observational and theoretical aspects of the cosmic web we refer to \cite{Shandarin:1989}, \cite{Bond:1996}, \cite{Weygaert:2008}, \cite{Aragon:2010a}, and \cite{Cautun:2014}.\\
\indent In this paper we present a framework to analytically quantify the cosmic web in terms of the statistics of these initial density fluctuations. In a Lagrangian description of gravitational structure growth, we see that caustics form where shell-crossing is occurring. We link the caustic features of the large-scale structure to singular points and curves in the initial conditions. We subsequently study the statistics of these singularities. We will concentrate on the linear Lagrangian approximation, known as the Zel'dovich 
approximation (see \cite{Zeldovich:1970}). We restrict our description to one and two spatial dimensions. It is straightforward to generalize the approach to spaces with higher 
dimensions. Moreover, we may forward it to the nonlinear regime by means of higher order corrections computed in Lagrangian effective field theory.

\section{The Zel'dovich approximation}
\noindent The gravitational collapse of small density fluctuations in an expanding universe can be modeled in multiple ways. In the Eulerian approach, we analyze the evolution of the (smoothed) density and velocity field. The resulting set of equations of motion is relatively concise, and leads to a reasonably accurate description of the mean flow field in a large fraction of space. However, due to its limitation to the mean velocity, the Eulerian formalism fails to describe the multistream structure of the matter distribution.\\

\indent In the Lagrangian approach, we assume that every point in space consists of a mass element. The evolution of the mass elements is described by means of a displacement field. The Zel'dovich approximation (\cite{Zeldovich:1970}) is the first order approximation of the displacement field of mass elements in the density field. It entails a product of a (universal) temporal factor and a spatial field of perturbations. The linear density growth factor $D_+(t)$ encapsulates all global cosmological information. The spatial component is represented by the gradient of the linearized velocity potential $\Psi(\textbf{q})$, and expresses the effect of the initial density fluctuation field. The Zel'dovich approximation thus translates into a ballistic motion,  expressed in terms of the displacement $\textbf{s}(\textbf{q},t)$ of a mass element with Lagrangian coordinate $\textbf{q}$, at time $t$, 
\begin{equation}
\textbf{s}(\textbf{q},t) = \textbf{x}(\textbf{q},t)-\textbf{q} = - D_+(t) \nabla_\textbf{q} \Psi(\textbf{q}),
\label{eq:zeldovich}
\end{equation}
with $\textbf{x}(\textbf{q},t)$ the Eulerian position of the mass element. Up to linear order the linearized velocity potential $\Psi(\textbf{q})$ is proportional to 
the linearly extrapolated gravitational potential at the current epoch $\phi_0(\textbf{q})$,
\begin{equation}
\Psi(\textbf{q}) = \frac{2}{3 \Omega_0 H_0^2}\phi_0(\textbf{q}),
\label{eq:tidefld}
\end{equation}
with Hubble constant $H_0$ and density parameter $\Omega_0$.\\
\indent The Lagrangian approach is aimed at the displacement of mass elements. We can express the density in terms of the displacement fields, i.e.
\begin{equation}
\rho(\textbf{x}',t)= \sum_{\textbf{q}\in A(\textbf{x}',t)} \left[\rho_u+\delta \rho_i(\textbf{q})\right]\left\|\frac{\partial \textbf{x}(\textbf{q},t)}{\partial \textbf{q}}\right\|^{-1},
\end{equation}
with $A(\textbf{x}',t)=\{\textbf{q}\ |\ \textbf{x}(\textbf{q},t)=\textbf{x}'\}$ the pre-image of the function $\textbf{x}(\textbf{q},t)$, the average initial density $\rho_u$, and the initial density fluctuation $\delta \rho_i(\textbf{q})$. The latter is formally equal to zero at the initial time $t=0$, because at the recombination epoch the average density $\rho_u$ dominates the fluctuations $\delta \rho_i(\textbf{q})$. For this reason the density evolution according to the Zel'dovich approximation is given as
\begin{equation}
\rho(\textbf{x}',t)=\sum_{\textbf{q}\in A(\textbf{x}',t)}\frac{\rho_u}{(1-D_+(t)\lambda_1(\textbf{q}))(1-D_+(t)\lambda_2(\textbf{q}))\dots (1-D_+(t) \lambda_d(\textbf{q}))}\label{eq:dens},
\end{equation}
with $\lambda_i$ the ordered eigenvalue fields, $\lambda_1(\textbf{q})\geq \lambda_2(\textbf{q}) \geq \dots \geq \lambda_d(\textbf{q})$ and corresponding eigenvector fields $\{\textbf{n}^{\lambda_i}(\textbf{q})\}$, of the deformation tensor
\begin{equation}
\psi_{ij}(\textbf{q})=\frac{\partial^2\Psi(\textbf{q})}{\partial q_i\partial q_j}, \qquad i,j=1,2,\dots,d,
\label{eq:deformtensor}
\end{equation}
or equivalently the tidal tensor (the Hessian of the gravitational potential $\phi_0$). From this, we see that the evolution of the density field is completely determined by the eigenvalue fields of the deformation tensor of the 
linear velocity potential and, through eqn.~\ref{eq:tidefld}, by the tidal field. 

\section{Caustics and catastrophe theory}
\begin{figure}[h]
\centering
\includegraphics[width = 0.99\textwidth]{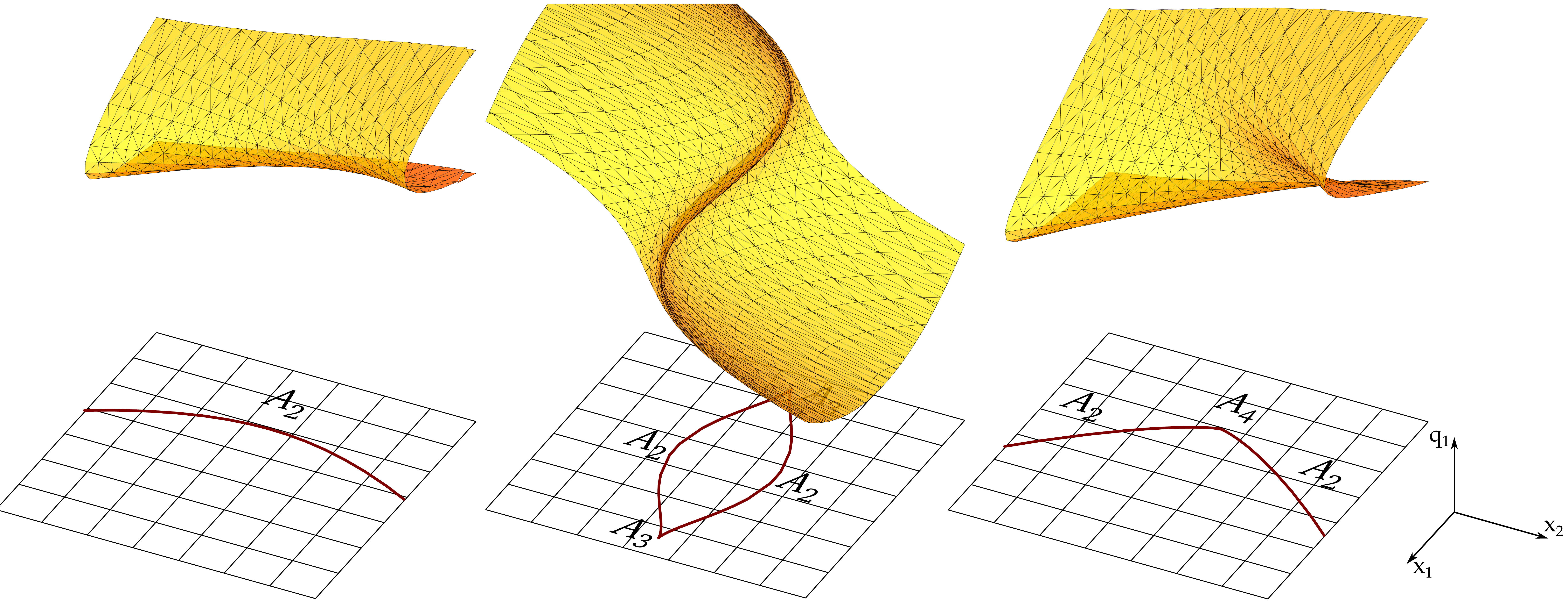} 
\caption{Visual illustrations of catastrophe classes $A_2$, $A_3$ and $A_4$ (\cite{Hidding:2014}). The surfaces represent a Lagrangian manifold in the 
phase space $(\textbf{q},\textbf{x})$ defined by the Lagrangian and Eulerian coordinates of mass elements,  
with on the horizontal axis 
the (final) Eulerian positions and on the vertical axis the (initial) Lagrangian positions. Note that we show only one of the two Lagrangian coordinates. The manifold is projected onto the Eulerian space. This results in several catastrophes. Left: an $A_2$ line catastrophe. Center: a Zel'dovich pancake singularity, consisting of two $A_2$ fold arcs and two $A_3$ cusp catastrophes. Right: a $A_4$ singularity.}
\label{fig:LC}
\end{figure}

\noindent The most prominent features of equation~\ref{eq:dens} are its poles. When at a given Lagrangian location $\textbf{q}$ at least one of the eigenvalue fields is positive, the density $\rho$ of that mass element can attains an infinite value at some time $t$. This circumstance occurs when mass elements temporarily accumulate in an infinitesimal volume. This defines a caustic. Physically we recognize these as the sites where matter streams start to cross (shell crossing). 
In mathematics, caustics have been extensively studied and are also known as catastrophes. We can identify them with the locations in configuration space (i.e, the space of final positions of the mass elements), where the projection of the Lagrangian phase-space
\footnote{We consider a phase space $(\textbf{q},\textbf{x})$ consisting of the (initial) Lagrangian $(\textbf{q})$ and (final) Eulerian positions $(\textbf{x})$ of the mass elements.}
 submanifold $(\textbf{q},\textbf{x})$ onto the configuration space leads to an infinite density by the accumulation of mass elements, i.e. where 
\begin{equation}
\left\|\frac{\partial {\bf x}}{\partial {\bf q}}\right\|=0
\end{equation} 
For a visual appreciation of this, we refer to the illustration in figure \ref{fig:LC}.

\cite{Arnold:1972} developed a classification of these (Lagrangian) catastrophes up to local coordinate transformations. Here we shortly summarize the 
main features of this classification. In the one-dimensional Zel'dovich approximation, poles can only (stably\footnote{Stable is understood in the sense that a small fluctuation of the Lagrangian manifold does not remove the caustic, but only shifts in in space and time. Note that the singularities of highest dimension only exist at a point in time. The $A_2$ and $A_3$ singularities in figure \ref{fig:LC} move in time while the $A_4$ singularity occurs in a transition of the singularity.}) occur in two manifestations, the so-called 
{\it fold} and {\it cusp} catastrophes. In the classification scheme of Arnol'd, these are denotes by $A_2$ and $A_3$. In two-dimensional space, there are two additional classes of (stable) catastrophes, to a total of four catastrophe classes. These are the {\it swallowtail} and {\it umbilic} catastrophes, denoted by $A_4$ and $D_4$. Finally, in 
three-dimensional space, we have a total of seven catastrophe classes. These include the additional $A_5$, $D_5$ and $E_5$ catastrophes. To obtain a visual 
impression of these catastrophes, figure~\ref{fig:LC} contains illustrations of the classes $A_2$, $A_3$ and $A_4$. The surfaces  
represent a Lagrangian manifold in the phase space $(\textbf{q},\textbf{x})$, defined by the Lagrangian and Eulerian coordinates of mass elements. 
The projection of these surfaces on to configuration space $\textbf{x}$ reveals itself in the spatial identity of the corresponding singularities. \\

\section{Zel'dovich and the cosmic skeleton}
\noindent Arnol'd, Shandarin and Zel'dovich (1982) linked the classification of catastrophes to the geometry of the deformation tensor eigenvalue fields in the one- and two-dimensional Zel'dovich approximation. In one dimension, the fold catastrophes correspond to level crossings of the eigenvalue field $\lambda$, whereas the cusp catastrophes 
correspond to maxima and minima of this eigenvalue field.

\begin{figure}[h]
\centering
 \includegraphics[width=0.99\textwidth]{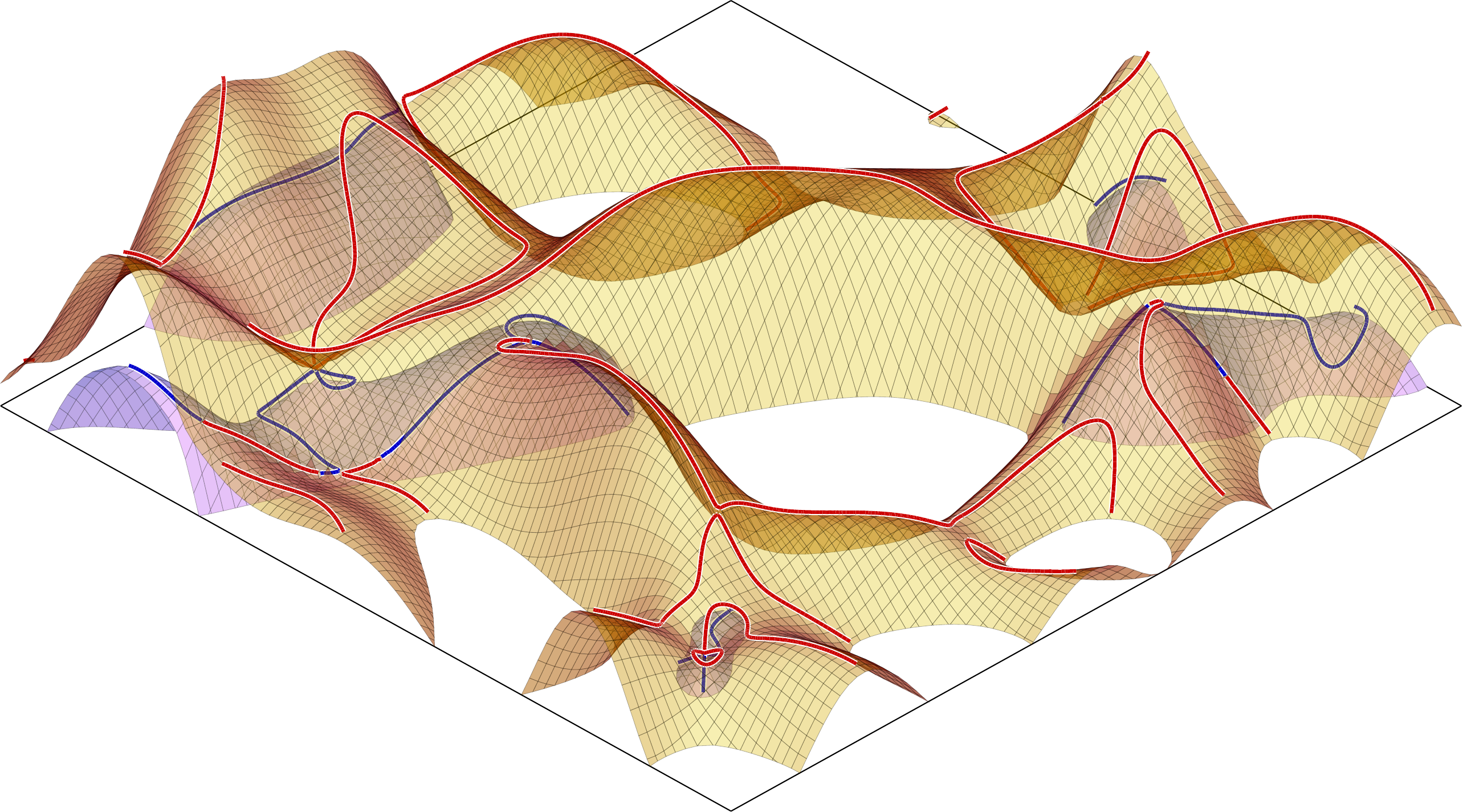} 
\caption{Iso-eigenvalue surfaces of an initial fluctuation field (in Lagrangian space). The yellow surface corresponds to the first eigenvalue $\lambda_1$ while the blue surface corresponds to the second eigenvalue field $\lambda_2$. Superimposed are a set of curves. The red and blue lines represent the $A_3$-lines corresponding to the first and second eigenvalue fields respectively. The eigenvalue fields of the (initial) Lagrangian density fluctuations completely determine the evolution in 
the Zel'dovich approximation. The caustic lines and points characterize the generation of singularities.}
\label{fig:Skeleton}
\end{figure}

\subsection{Catastrophes in the 2D Zel'dovich formalism} 

\noindent On the basis of catastrophe theory, we may classify the caustics that arise in a cosmic density field that evolves according to the 
Zel'dovich approximation. The classification is based on the spatial characteristics of the field of the first and second eigenvalues 
$\lambda_1$ and $\lambda_2$ of the deformation tensor $\psi_{ij}$ (eqn~\ref{eq:deformtensor}). Following \cite{Arnold:1982}, we may then identify 
the corresponding set of $A_3$-lines, and $A_3$, $A_4$ and $D_4$ points (see figure \ref{fig:Skeleton}, from \cite{Hidding:2014}, for an illustration). 
These caustic points and curves describe the regions where infinite densities arise in the Zel'dovich approximation.\\

\noindent In two dimensions, the fold catastrophes correspond to isocontours of the first or second eigenvalue fields $\lambda_1$ or $\lambda_2$, known as $A_2$-lines defined as $A_2^i(\lambda)=\{ \textbf{q}\ |\  \lambda_i(\textbf{q})=\lambda\}$ (see figure~\ref{fig:Skeleton}). The cusp catastrophes are located on the $A_2$-lines at which the eigenvector ${\textbf n}^{\lambda_i}$ is orthogonal to the gradient of the corresponding eigenvalue field,
\begin{equation}
{\textbf n}^{\lambda_i}(\textbf{q}) \cdot \nabla \lambda_i(\textbf{q})= 0,  \qquad i=1,2.
\end{equation}
Generically, these points form a piecewise smooth curve, known as the $A_3$-line defined as $A_3^i=\{ \textbf{q}\ |\ {\textbf n}^{\lambda_i}(\textbf{q}) \cdot \nabla \lambda_i(\textbf{q})= 0\}$
 (see figure \ref{fig:Skeleton}).\\
\indent Note that the caustic points on the $A_3$- line assume their $A_3$ singularity state at different times: starting at a maximum on the $A_3$ line, the location of 
the cusps moves along the line towards lower values of the eigenvalue $\lambda_i$. Hence, we observe that the maxima and saddle points of the eigenvalue field are special cusp catastrophes. In 
the Zel'dovich approximation, the maxima mark the points at which the first infinite densities 
emerge. Subsequently, as time proceeds, the caustics at the maxima become Zel'dovich pancakes consisting of two $A_2$ fold arcs (see figure \ref{fig:LC}). 
The cusps at the tips of the pancake are defined by the corresponding points on the $A_3$-line. Within this context, we observe the merging of 
two pancakes at the saddle points in the eigenvalue field. 

The $A_4$ and $D_4$ catastrophes occur at the singularities of the $A_3$-line. The swallowtail catastrophe occurs when the tangent of the 
$A_3$-line becomes parallel to the isocontour of the corresponding eigenvalue field. The umbilic catastrophes occur in points where the two eigenvalue fields coincide. 
We refer to \cite{Hidding:2014} for a more detailed study of the geometric and dynamic nature of the catastrophes in the two-dimensional Zel'dovich approximation.

\begin{figure}[h]
\centering
 \includegraphics[ width=0.99\textwidth]{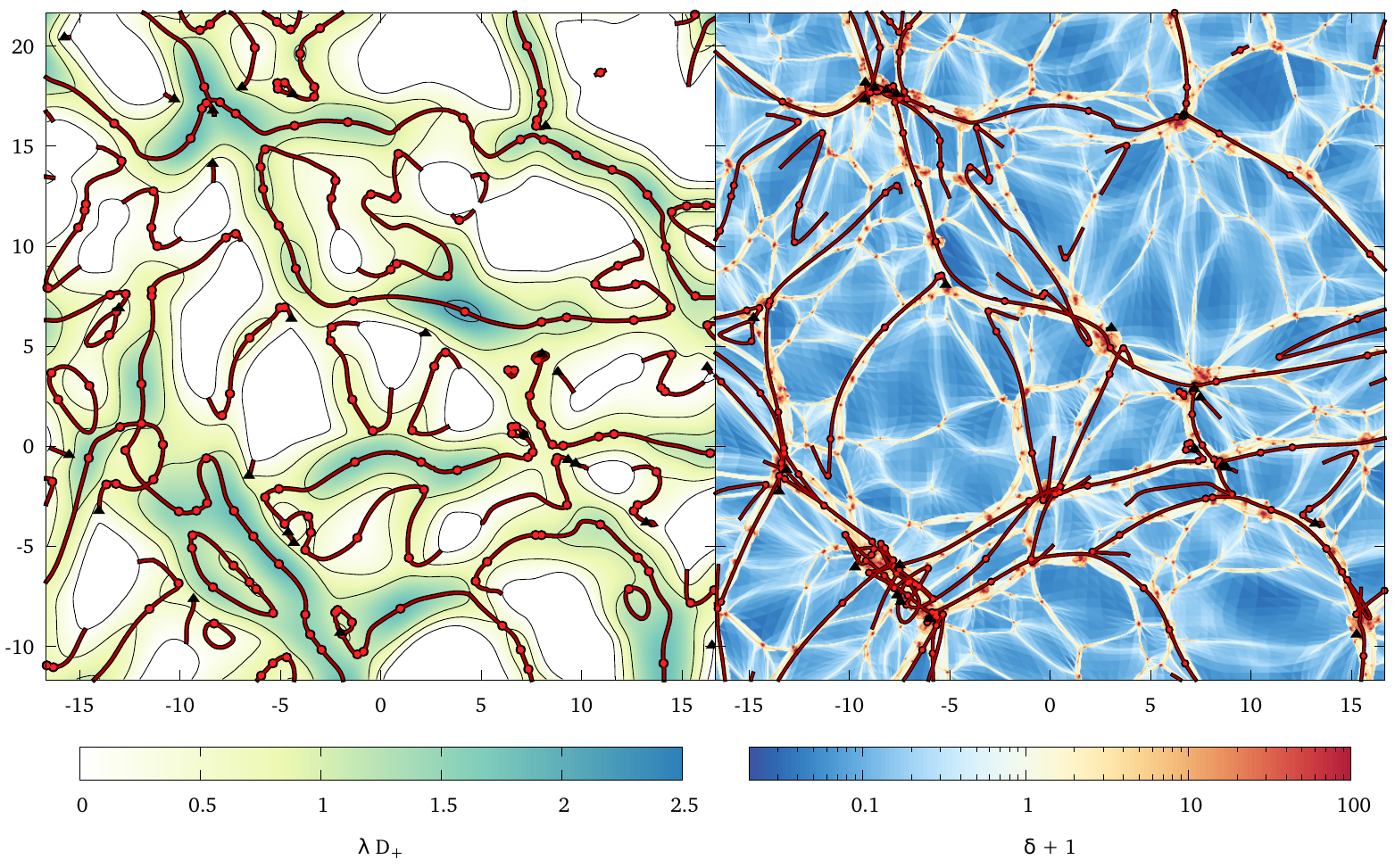} 
\caption{
Spatial distribution of singularities in Lagrangian and Eulerian Cosmic Web. The figure compares the spine of the cosmic web with the mass 
distribution in a 2-D $N$-body simulation. Left panel: initial field of density fluctuations and the skeleton of identified singularities/catastrophes. Right panel: 
density field of an evolved 2D cosmological $N$-body simulation, in which the Lagrangian skeleton of singularities is mapped by means of the Zel'dovich 
approximation. From Feldbrugge et al. 2014.}
\label{fig:Nbody}
\end{figure}

\subsection{Catastrophes and the cosmic skeleton}
\indent Although the Zel'dovich approximation is only accurate in the linear and quasi-nonlinear regime, the emerging caustics turn 
out to be manifest themselves in the present-day nonlinearly evolved large-scale structure. They are a proxy for the cosmic skeleton 
of the evolving weblike mass distribution. This can be directly appreciated from figure \ref{fig:Nbody}. 
It compares a two-dimensional dark matter $N$-body simulation with the distribution of caustic curves and points inferred from 
the Zel'dovich approximation. 

The initial Gaussian random density field has a power-law power spectrum. The left panel of figure~\ref{fig:Nbody} contains an 
isocontour map of the first eigenvalue field $\lambda_1$. Superimposed on this map we find the corresponding $A_3$-lines and 
$A_3$ and $D_4$ points. The depicted caustics are the ones obtained from the initial density field filtered with a Gaussian 
kernel, such that the fluctuation amplitude in the linearly extrapolated density slightly exceeds unity. In this manner, 
the extracted skeleton corresponds to structures that have just entered the collapse phase. 

The mass distribution in the right panel is the result of the gravitational growth of the initial Gaussian random field.
To follow this, we evolved the initial density and velocity field by a two-dimensional dark matter $N$-body simulation. 
The resulting density field is marked by a weblike pattern of clusters, filaments and voids. Superimposed on the density map 
are the $A_3$-lines, $A_3$, $A_4$ and $D_4$ points, mapped from their Lagrangian towards their Eulerian location by
means of the Zel'dovich approximation (eqn.~\ref{eq:zeldovich}).\\

\noindent Comparison between both panels reveals a close match between the weblike structure in the density field and the spatial distribution of the 
catastrophe points and lines. The point catastrophes are mostly located in the dense cluster nodes, while the $A_3$-lines are found to closely trace 
the filaments. It demonstrates the fact the spine of the cosmic web is already outlined by the catastrophes in the Gaussian initial conditions in 
the Lagrangian volume. In this process, the main structure remains largely intact. It forms a justification for describing the cosmic web in terms 
of the defining caustic singularities, as may be inferred from detailed studies of the evolving mass and halo distribution in and around the cosmic web 
(see e.g. \cite{Aragon:2010a}, \cite{Cautun:2014} and \cite{Robles:2014}). 

It is also good to emphasize the dynamical nature of this definition of the spine of the cosmic web. In this, it differs from 
the skeleton of the cosmic web identified on the basis of structural and topological aspects of the density field (e.g. \cite{Novikov:2006}, 
\cite{Aragon:2007}, \cite{Sousbie:2008}, \cite{Aragon:2010b}, \cite{Sousbie:2011} and \cite{Cautun:2013}). On the other hand, it establishes 
a close link to the reecent studies by \cite{Shandarin:2012}, \cite{Abel:2012} and \cite{Neyrinck:2012}, who assessed the phase-space 
structure of the cosmic mass distribution to identify the various morphological elements of the cosmic web. 

\section{Statistics of caustics}
\noindent As we argued in the previous sections, the catastrophes of the one- and two-dimensional Zel'dovich approximation can be linked to local properties of the deformation tensor eigenvalue fields in the (initial) Lagrangian density field. We assume that this density field closely resembles a Gaussian random field. So far, no counter-evidence for this (i.e., non-Gaussianities) has been found in observations of the cosmic microwave background radiation field. Moreover, the assumption follows naturally from both inflation theory and the central limit theorem.\\
\indent Gaussian random fields have been extensively studied. \cite{Doroshkevich:1970} was among the first to study the eigenvalue field of Gaussian random fields. \cite{BBKS} and \cite{Adler} studied the statistics of Gaussian random fields near critical points (also see \cite{Weygaert:1996,Catelan:2001,Desjacques:2008,Rossi:2012}).\\
\indent A Gaussian random field is completely characterized by the second order moment of the fluctuation field, i.e., by its autocorrelation function or, equivalently, its power spectrum. For any finite number of points $\textbf{q}_1,\textbf{q}_2,\dots,\textbf{q}_n\in\mathbb{R}^d$, the probability distribution that the density field $f:\mathbb{R}^d\to \mathbb{R}$ assumes the values $f_i\in\mathbb{R}$ in $f(\textbf{q}_i)$ for $i=1,2,\dots,n$ is given by
\begin{equation}
p(f(t_1),\dots,f(t_n))=\frac{\exp\left[-\frac{1}{2} \sum_{i,j} f(t_i)(M^{-1})_{ij}f(t_j)\right]}{[(2\pi)^n\det M]^{1/2}}, \quad
\end{equation}
in which the matrix elements $M_{ij}$ express the spatial 2pt correlation of the field $f(\textbf{q})$, 
\begin{equation}
M_{ij}=\langle f(\textbf{q}_i)f(\textbf{q}_j)\rangle.\label{eq:GRF}
\end{equation}
Using equation \ref{eq:GRF}, we can now calculate statistical properties of the caustics in the one- and two-dimensional Zel'dovich approximation.\\\\
\noindent Here we evaluate the number density of critical points - maxima, minima and saddle points - and catastrophe points, as well as the average line length of catastrophe lines in (initial) Lagrangian space. It is straightforward to extend this description to the curves and points in 
Eulerian space. 

In order to calculate the density of points in random fields we use Rice's formula (\cite{Rice:1944,Rice:1945}). The number density of points $\textbf{q}$ for which a function $f=(f_1,f_2):\mathbb{R}^2\to \mathbb{R}^2$ takes a value $\textbf{y}=(y_1,y_2)\in\mathbb{R}^2$ is given by
\begin{equation}
\mathcal{N}=\left\langle \delta^{(2)}(f-\textbf{y})\det(f_{i,j})\right\rangle
=\int p(f_1=y_1,f_2=y_2,f_{i,j})\det(f_{i,j})\mathrm{d}f_{1,1}\dots \mathrm{d}f_{2,2},
\end{equation}
with $f_{i,j}=\frac{\partial f_i}{\partial q_j}$.
By using the properties of the eigenvalue fields at the caustics, the density of the $A_2$ fold and $A_3$ cusp catastrophes in the one-dimensional case can be expressed as
\begin{eqnarray}
\mathcal{N}_{A_2}(\alpha)&\,=\,&\int |\lambda_{1,1}|\,p(\lambda_1=\alpha,\lambda_{1,1})\mathrm{d}\lambda_{1,1},
\label{eq:1}\\
\mathcal{N}_{A_3}(\alpha)&\,=\,&\int |\lambda_{1,11}|\, p(\lambda_1=\alpha,\lambda_{1,1}=0,\lambda_{1,11})\mathrm{d}\lambda_{1,11}\label{eq:2}\,.
\end{eqnarray}
Note that in equation \ref{eq:1} we integrate over $\mathbb{R}$ whereas in equation \ref{eq:2} we integrate over $(-\infty,0)$ for the maxima and $(0,\infty)$ for the minima of $\lambda_1$. In the two-dimensional Zel'dovich approximation, the density of $D_4$ points can be computed by evaluating the integral
\begin{equation}
\mathcal{N}_{D_4}(\alpha)=\int |\lambda_{1,1} \lambda_{2,2}-\lambda_{1,2}\lambda_{2,1}| \,
p(\lambda_1=\alpha,\lambda_2=\alpha,\lambda_{1,1},\lambda_{1,2},\lambda_{2,1},\lambda_{2,2}) \mathrm{d}\lambda_{1,1}\mathrm{d}\lambda_{1,2}\mathrm{d}\lambda_{2,1}\mathrm{d}\lambda_{2,2}.
\end{equation}
The density of $A_3$ and $A_4$ points in the two-dimensional case are obtained in an analogous fashion.\\
\indent For the $A_3$-lines in the two-dimensional case, we study the curve length density, by adapting the statistical analysis of \cite{Longuet:1957}. The average length of the iso-contour of a function $f:\mathbb{R}^2\to \mathbb{R}$ at level $y$ is given by
\begin{equation}
\mathcal{L}(y)=\left\langle \delta^{(1)}(f-y)\sqrt{f_{,1}^2+f_{,2}^2} \right\rangle=\int p(f=y,f_{,1},f_{,2})\sqrt{f_{,1}^2+f_{,2}^2}\ \mathrm{d}f_{,1}\mathrm{d}f_{,2}.
\end{equation}
By using applying the eigenvalue conditions of the $A_3$-lines, we obtain the differential $A_3$-line length with respect to the first eigenvalue field $\lambda_1$, 
\begin{equation}
\mathcal{L}_{A_3}(\lambda_1)=\pi \int \sqrt{\lambda_{1,11}^2+\lambda_{1,12}^2}\ \,
p(\lambda_1,\lambda_2,\lambda_{1,1}=0,\lambda_{1,11},\lambda_{1,12})
 (\lambda_1-\lambda_2)\mathrm{d}\lambda_{1,11}\mathrm{d}\lambda_{1,12}\mathrm{d}\lambda_{2}.
\end{equation}
Other local properties, such as the curvature or the correlation function between caustics, can be determined analogously.

\section{Conclusion}
\noindent Here we have presented a formalism to describe the spatial statistics of caustics in the one- and two-dimensional Zel'dovich approximation, for a given power spectrum of the initial random Gaussian density field. The visual comparison of the spatial distribution of these caustics with the pattern of the mass distribution in 
$N$-body simulations demonstrates that the caustics define the spine of the cosmic web. It reflects the strong correspondence between catastrophe lines and 
points and the emerging weblike structures in the cosmic mass distribution. In other words, the skeleton of the cosmic web appears to be defined by the 
spatial properties of the tidal force and deformation field in the initial Gaussian mass distribution. 

It is straightforward to extend the one- and two-dimensional formalism presented here to three dimensions. Moreover, currently we are looking into how to 
extend the formalism to more advanced stages of dynamical evolution, using Lagrangian effective field theory.


\begin{thebibliography}{}
\bibitem[Abel \etal\, (2012)]{Abel:2012}
{Abel, T. , Hahn, O., Kaehler, R.}, 2012,
\textit{MNRAS}, 427, 61
\bibitem[Adler (1981)]{Adler}
{Adler, R.J.}, 1981, 
\textit{The Geometry of Random Fields}, Wiley
\bibitem[Adler \& Taylor (2007)]{Adler:2007}
{Adler, R.J., Taylor, J.E.}, 2007,
\textit{Random Fields and Geometry}, Springer
\bibitem[Arag\'on-Calvo \etal\, (2007)]{Aragon:2007}
{Arag\'on-Calvo, M.A., Jones, B.J.T, van de Weygaert,R., van der Hulst, J.M.}, 2007,
\textit{Astron. Astrophys}, 474, 315
\bibitem[Arag\'on-Calvo \etal\, (2010a)]{Aragon:2010a}
{Arag\'on-Calvo, M.A., van de Weygaert, R. , Jones, B.J.T.}, 2010,
\textit{MNRAS}, 408, 2163
\bibitem[Arag\'on-Calvo \etal\, (2010b)]{Aragon:2010b}
{Arag\'on-Calvo, M.A., Platen, E., van de Weygaert,R., Szalay, A.S.}, 2010,
\textit{Astrophys. J.},  723, 364 
\bibitem[Arnol'd (1972)]{Arnold:1972}
{Arnol'd V.I.}, 1972, 
\textit{Funct. Anal. and its Appl.}, 6, 254
\bibitem[Arnol'd \etal\, (1982)]{Arnold:1982}
{Arnol'd, V.I., Shandarin, S.F., Zeldovich, Ia.B.}, 1982,
\textit{Geophys. and Astrophys. Fluid Dynamics}, 20, no. 1-2, 111
\bibitem[Bardeen \etal\, (1986)]{BBKS}
{Bardeen, J.M., Bond, J.R., Kaiser, N., Szalay, A.S.}, 1986,
\textit{Astrophys. J.}, 304, 15
\bibitem[Bond, Kofman \& Pogosyan (1996)]{Bond:1996}
{Bond, J.R., Kofman, L., Pogosyan, D.}, 1996,
\textit{Nature}, 380, 603
\bibitem[Catelan \& Porciani (2001)]{Catelan:2001}
{Catelan, P., Porciani, C.}, 2001,
\textit{MNRAS}, 323, 713
\bibitem[Cautun \etal\, (2013)]{Cautun:2013}
{Cautun, M., van de Weygaert, R. , Jones}, 2013,
\textit{MNRAS}, 429, 1286
\bibitem[Cautun \etal\, (2014)]{Cautun:2014}
{Cautun, M., van de Weygaert, R. , Jones, B.J.T., Frenk, C.S.}, 2014,
\textit{MNRAS}, 441, 2923
\bibitem[Desjacques \& Smith (2008)]{Desjacques:2008}
{Desjacques, V., Smith, R.E.}, 2008,
\textit{Phys. Rev. D}, 78, 023527
\bibitem[Doroshkevich (1970)]{Doroshkevich:1970}
{Doroshkevich, A.G.}, 1970,
\textit{Astrophysics}, 6, Issue 4, 320
\bibitem[Hidding \etal\, (2014)]{Hidding:2014}
{Hidding, J., Shandarin, S.F., van de Weygaert, R.}, 2014,
\textit{MNRAS}, Vol. 437, , 3442
\bibitem[Longuet-Higgins (1957)]{Longuet:1957}
{Longuet-Higgins, M.S.}, 1957,
\textit{Philosophical Trans. of the Royal Society of London} Vol. 249, 321
\bibitem[Neyrinck (2012)]{Neyrinck:2012}
{Neyrinck, M.}, 2012, 
\textit{MNRAS}, 427, 494
\bibitem[Novikov \etal\, (2006)]{Novikov:2006}
{Novikov, D., Colombi, S., Dor\'e, O.}, 2006, 
\textit{MNRAS}, 366, 1201
\bibitem[Rice (1944)]{Rice:1944}
{Rice, S.O.}, 1944,
\textit{Bell Systems Tech. J.}, 23, 282
\bibitem[Rice (1945)]{Rice:1945}
{Rice, S.O.}, 1945,
\textit{Bell Systems Tech. J.}, 24, 46
\bibitem[Robles (2014)]{Robles:2014}
{Robles S., Dom\'{\i}nguez-Tenreiro R., O\~norbe J., Mart\'{\i}nez-Serrano F.}, 2014, 
\textit{NMRAS}, subm.
\bibitem[Rossi (2012)]{Rossi:2012}
{Rossi G.}, 2012,
\textit{MNRAS}, 421, 296
\bibitem[Shandarin \etal\, (2012)]{Shandarin:2012}
{Shandarin, S.F., Habib, S., Heitmann, K.}, 2012,
\textit{Phys. Rev. D}, 85, 083005
\bibitem[Sousbie (2008)]{Sousbie:2008}
{Sousbie, T., Pichon, C., Colombi, S., Novikov, D., Pogosyan, D.}, 2008,
\textit{MNRAS}, 383, 1655
\bibitem[Sousbie (2011)]{Sousbie:2011}
{Sousbie, T.}, 2011,
\textit{MNRAS}, 414, 350
\bibitem[Shandarin \& Zel'dovich (1989)]{Shandarin:1989}
{Shandarin, S.F., Zel'dovich, Ia.B.}, 1989,
\textit{Rev. Mod. Phys.}, 61, 185
\bibitem[van de Weygaert \& Bertschinger (1996)]{Weygaert:1996}
{van de Weygaert, R., Bertschinger, E.}, 1996,
\textit{MNRAS}, 281, 84
\bibitem[van de Weygaert \& Bond (2008)]{Weygaert:2008}
{van de Weygaert, R. , Bond, J.R.}, 2008,
\textit{Lecture Notes in Physics}, 740, 335
\bibitem[Zel'dovich (1970)]{Zeldovich:1970}
{Zel'dovich Ia.B.}, 1970, 
\textit{Astron. Astrophys.}, 5, 84
\end{thebibliography}
\end{document}